\documentclass[aip,jcp,amsmath,preprint,reprint]{revtex4-1}
\usepackage{dcolumn}
\usepackage{hyperref}

\DeclareMathOperator{\tr}{tr}
\renewcommand{\vec}[1]{{\rm\bf #1}}

\newcommand{\de}{{\rm d}}
\newcommand{\te}{{\mathsf{T}}}

\begin{document}

\title{Optimization of atomic density-fitting basis functions \\ for molecular two-electron integral approximations}

\author{Dimitri N. Laikov}
\email[]{laikov@rad.chem.msu.ru}
\homepage[]{http://rad.chem.msu.ru/~laikov/}
\affiliation{Chemistry Department, Moscow State University,
119991 Moscow, Russia}

\date{\today}

\begin{abstract}
A general procedure for the optimization of atomic density-fitting basis functions
is designed with the balance between accuracy and numerical stability in mind.
Given one-electron wavefunctions and energies,
weights are assigned to the product densities,
modeling their contribution to the exchange and second-order correlation energy,
and a simple weighted error measure is minimized.
Generally-contracted Gaussian auxiliary basis sets are optimized
to match the wavefunction basis sets
[D. N. Laikov, Theor. Chem. Acc. \textbf{138}, 40 (2019)]
for all 102 elements
in a scalar-relativistic approximation
[D. N. Laikov, J. Chem. Phys. \textbf{150}, 061102 (2019)].
\end{abstract}

\maketitle

\section{Introduction}

Approximations of molecular two-electron repulsion integrals
by fitting the underlying two-center product densities
to sums of atom-centered basis functions
have a long
history~\cite{F57,SP62,N69,BB71,W73,BER73,SF75,DCS79,A88,AW92,VAF93,FFK93,RL94,JH96,FKHD97,WH97,GSVB98,HW00,D00,WMK03,PWMK04,SWLM04,GKH05,GGTFH05,JSGH05,DPAL15,HPM12,MKHLARP13}
and are now in widespread use,
even three-electron~\cite{TM03,MM04,WM14} integrals of explicitly-correlated methods
can be decomposed this way, to speed up the overall workflow.
The accuracy then depends on the size and quality
of the auxiliary density-fitting basis set
that can be built in many ways.
An on-the-fly generation~\cite{BL77,TI95,ALB07}
of linearly independent combinations of one-center products
is easy but leads to a slower evaluation of three-center integrals.
Dedicated sets of simple atomic functions
have been carefully prepared,
by some rules of thumb and optimization techniques,
for Coulomb~\cite{GSAW92,ETOHA95,LB03,W06}, exchange~\cite{W02,W07},
and second-order correlation~\cite{WHPA98,WKH02,H05,HHHK07} energy components,
and all three together toward an automatic generation
of ``universal''~\cite{SAN17} solutions.

We have also developed our own general procedure
for the optimization of atomic density-fitting basis sets
for molecular two-electron integral approximation,
for all two-electron energy components,
which is new in its kind and even more simple than the others~\cite{WHPA98,SAN17},
while giving a good balance between accuracy and numerical stability.
Its older version has been used since 15 years ago
to match and follow our first wavefunction basis work~\cite{L05},
but was never published on psychological~\cite{U74} grounds,
although those primary and auxiliary basis sets
have since been used in hundreds of
published~\cite{SS06,SSV07,BBN07,MABTL07,SSMH08,SKB08,SS08,DZLGKGG09,PSS10,MWEKS10,TSF10,ZRKB11,DSKVFKBBBSHG11,BPSKD11,SLBF11,AICDP11,
DTBBTZ12,MRPK12,PKTKLNK12,EPRCDPE13,AHNLGCMMCYOS13,MOJ13,BPSK13,BSMMSMKIS14,UGKMBAU14,LBK14,MLGAKCBK14,EGVNGBE14,
SLMCC15,UBBGMKATKU15,VBMFGSVA15,GSSZK17,MBMOAMQCDGB18}
studies.
Overcoming the barrier, we now reveal our procedure
and apply it to our newer basis sets~\cite{L19b}
within a scalar-relativistic approximation~\cite{L19a}
covering all 102 elements hydrogen through nobelium.

\section{Theory}

We take as input
a canonical set of $N$ atomic one-electron wavefunctions~\cite{L19b}
$\{\phi_i(\vec{r})\}$ and $\{\phi_a(\vec{r})\}$
with energies $\{\epsilon_i\}$ and $\{\epsilon_a\}$,
of which $N_\mathrm{o}$ are occupied,
labeled with $i,j=1,\dots,N_\mathrm{o}$,
and $a,b=N_\mathrm{o}+1,\dots,N$.

From order-of-magnitude arguments,
we set the weights for the product densities,
\begin{eqnarray}
w_{ij} &=& 1, \\
\label{eq:wai}
w_{ai} &=& \beta \frac{\bar{K}_{ai}}{\epsilon_a - \epsilon_i}, \\
\label{eq:g}
w_{ab} &=& \gamma \sum_i w_{ai} w_{bi},
\end{eqnarray}
where the exchange integrals
\begin{equation}
\label{eq:kai}
K_{ai} = \left(\phi_a \phi_i \left|\right.\! \phi_a \phi_i \right)
\end{equation}
are spherically-averaged,
\begin{equation}
\label{eq:kw}
\bar{K}_{ai} =
\frac{\sum_{bj} \delta_{l_a l_b} \delta_{l_i l_j} \delta_{n_a n_b} \delta_{n_i n_j} K_{bj}}
     {(2l_a + 1)(2l_i + 1)} ,
\end{equation}
the charge density notation
\begin{equation}
\left(p \left| \right.\! q \right)
= \int
\frac{p(\vec{r}_1) q(\vec{r}_2)}
     {\left|\vec{r}_1 - \vec{r}_2\right|}
 \de^3 \vec{r}_1 \de^3 \vec{r}_2
\end{equation}
being used throughout.
The constants $\beta$ and $\gamma$ are quite arbitrary,
while the first term of the weighted sum
\begin{equation}
\label{eq:e0}
\mathcal{E}_0 = \tfrac12 \sum_{ij} w_{ij} K_{ij} + \sum_{ai} w_{ai} K_{ai} + \tfrac12 \sum_{ab} w_{ab} K_{ab}
\end{equation}
is the exchange energy,
we set $\beta=\tfrac12$ making the second term
\textit{twice} the opposite-spin second-order correlation energy
in the special case of helium atom and the smallest \texttt{L1}~\cite{L19b} basis set,
the gradient then has, in this linear model, both components \textit{equally} weighted in the optimization.
Either $\gamma = 0$ or $\gamma = 1$ will be taken to see what happens.

We now label the set of $N^2$ densities and weights
\begin{eqnarray}
\label{eq:pk}
\rho_k(\vec{r}) &=& \phi^\dagger_{k\bmod N}(\vec{r})\cdot \phi_{\lfloor k/N \rfloor}(\vec{r}), \\
w_k &=& w_{k\bmod N, \lfloor k/N \rfloor},
\end{eqnarray}
with one linear index $k = 1,\dots,N^2$.

A set of $M$ primitive density-fitting basis functions $\{\varrho_\mu(\vec{r},\alpha_\mu)\}$,
$\mu,\nu = 1,\dots,M$,
will be contracted to $M_\mathrm{o}$ linear combinations $\{\varrho_u(\vec{r})\}$,
$u = 1,\dots,M_\mathrm{o}$,
\begin{equation}
\varrho_u(\vec{r}) = \sum_\mu \varrho_\mu(\vec{r},\alpha_\mu) \, B_{\mu u},
\end{equation}
optimizing both the linear coefficients $\{B_{\mu u}\}$
and the nonlinear parameters $\{\alpha_\mu\}$ (such as exponents)
for the best approximation of the densities
\begin{equation}
\varrho_k(\vec{r}) = \sum_u \varrho_u(\vec{r}) \, C_{uk}
\approx \rho_k(\vec{r})
\end{equation}
by minimizing the weighted error measure
\begin{equation}
\label{eq:e}
\mathcal{E} = \tfrac12
\sum_k \left(\varrho_k - \rho_k \left| \right.\! \varrho_k - \rho_k \right) w_k .
\end{equation}
The solution for $\{C_{uk}\}$ is straightforward,
\begin{equation}
\label{eq:c}
\vec{C} = \left(\vec{B}^\te \vec{V} \vec{B} \right)^{-1} \vec{B}^\te \vec{R} ,
\end{equation}
given in matrix form with the integrals
\begin{eqnarray}
R_{\mu k}  &=& \left(\varrho_\mu \left| \right.\! \rho_k \right) , \\
V_{\mu\nu} &=& \left(\varrho_\mu \left| \right.\! \varrho_\nu \right) ,
\end{eqnarray}
so Eq.~(\ref{eq:e}), with the help of Eqs.~(\ref{eq:e0}) and~(\ref{eq:c}), becomes
\begin{equation}
\label{eq:eb}
\mathcal{E} = \mathcal{E}_0
 - \tfrac12 \tr \left\{
\vec{W} \vec{R}^\te \vec{B} \left(\vec{B}^\te \vec{V} \vec{B} \right)^{-1} \vec{B}^\te \vec{R}
\right\} ,
\end{equation}
\begin{equation}
W_{kk'} = \delta_{kk'} w_k .
\end{equation}
Since the problem is invariant to such transformations,
we seek an orthonormal $\vec{B}$ with Coulomb metric
\begin{equation}
\vec{B}^\te \vec{V} \vec{B} = \vec{1},
\end{equation}
then the minimization of Eq.~(\ref{eq:eb})
leads to the solution of eigenvalue problem
\begin{equation}
\label{eq:b}
\left( \vec{R} \vec{W} \vec{R}^\te - \varepsilon_u \vec{V} \right) \vec{B}_u = 0,
\end{equation}
taking the $M_\mathrm{o}$ column eigenvectors $\{\vec{B}_u\}$
with the highest eigenvalues $\{\varepsilon_u \}$.

The nonlinear parameters can be optimized using the first derivatives
\begin{equation}
\frac{\partial\mathcal{E}}{\partial\alpha_\mu} =
 \tr \left\{ \vec{W} \vec{C}^\te \vec{B}^\te \left(
 \tfrac12 \frac{\partial\vec{V}}{\partial\alpha_\mu} \vec{B} \vec{C}
 - \frac{\partial\vec{R}}{\partial\alpha_\mu}
\right) \right\} ,
\end{equation}
and, if needed, the lengthy but straightforward second derivatives.
As before~\cite{L19b}, we parametrize the exponents
\begin{equation}
\alpha_\mu = \exp\left(x_1 + \sum\limits_{\nu=2}^\mu \sqrt{x_0^2 + x_\nu^2} \right),
\end{equation}
taking $\{x_\mu\}$, $\mu \ge 1$, as the optimization variables,
and setting $x_0 = (\ln 2)/4$ to bound $\alpha_{\mu+1} /\alpha_\mu \ge 2^{1/4}$
against collapses $\alpha_{\mu+1} \to \alpha_\mu$.

After the full optimization,
we make an orthogonal transformation of coefficients $\{\vec{B}_u\}$
to zero out the charge moments of all but one function $\varrho_u(\vec{r})$
within each $(l,m)$ angular symmetry block
and, furthermore, to zero out the triangular blocks of coefficients
of the most diffuse functions $\varrho_\mu(\vec{r})$ ---
helping make the three-index integral supermatrices much sparser
in calculations on large polyatomic systems.

For molecules, we replace the exact inversion of an often ill-conditioned matrix
with the approximation~\cite{ETOHA95}
\begin{equation}
\vec{V}^{-1} \approx (\vec{V} + \upsilon)^{-1} \left( 2 - \vec{V} (\vec{V} + \upsilon)^{-1} \right)
\end{equation}
and handle the shifted matrix $(\vec{V} + \upsilon)^{-1}$
by the Householder transformation~\cite{H58}, thanks to its good parallelizability,
followed by a simple solution of the tridiagonal linear system.
We find $\upsilon = 2^{-16}$ to work well and use it in the tests that follow.

All this said, we understand the limits of our method:
at best, we would have the set $\{\varrho_u(\vec{r})\}$ spanning
all atomic one-center products $\{\rho_k(\vec{r})\}$,
but it would then fit the two-center products in molecules
only to some finite (luckily good enough) accuracy
that cannot be improved any further.
A definitive solution would be to work with both one- and two-center products
at a full range of distances, minimizing the sum
\begin{equation}
\mathcal{E} = \sum\limits_I \mathcal{E}_I
+ w_2 \sum\limits_{I<J} 4\pi \int\limits_{R_{IJ}}^\infty \mathcal{E}_{IJ}(R) \; R^2 \de R
\end{equation}
of the atomic errors $\mathcal{E}_I$ as in Eq.~(\ref{eq:e})
for each element $I$ (such as $I=1,\dots,102$),
and the diatomic errors $\mathcal{E}_{IJ}(R)$ where
products of wavefunctions in Eq.~(\ref{eq:pk})
are taken for two elements $I$ and $J$ on two centers at distance $R$
and the same is done for the weights in Eqs.~(\ref{eq:wai}), (\ref{eq:kai}),
and~(\ref{eq:kw}), while the density-fitting functions are used at both centers.
The starting distances can be chosen as averages of radial extents
of occupied wavefunctions,
\begin{eqnarray}
R_{IJ} &=& \tfrac12 \left( R_I + R_J \right), \\
R_I &=& \max_i \left<\phi_{i,I} \left| r \right|\! \phi_{i,I} \right> .
\end{eqnarray}
The diatomic weight $w_2$ seems to be arbitrary and can be experimented with.
A periodic-table-wide optimization of this kind would be computationally intensive
and would also need extended-precision arithmetic to cope with ill-conditioned problems ---
we do not now, but may one day, undertake it.

\section{Calculations}

We have implemented our methodology into our atomic electronic structure program~\cite{L19b}
making use of spherical symmetry and working in up to 256-bit precision,
and apply it to a full collection of atomic wavefunctions~\cite{L19b}
from our work on a scalar-relativistic approximation~\cite{L19a}.

Having done all the work twice
with either $\gamma = 1$ or $\gamma = 0$ in Eq.~(\ref{eq:g}),
we have seen only small changes in test results for molecules,
so we set $\gamma = 0$ henceforth.

The number of functions in a density-fitting basis set
should be carefully chosen ---
we have used our natural intelligence
to find regularities,
among a series of atoms,
in the distribution of eigenvalues of Eq.~(\ref{eq:b})
for each angular symmetry,
aligning them rowwise and columnwise in the periodic table
until an acceptable solution has been found ---
it is shown in Table~\ref{tab:n} where the sizes of wavefunction and density sets are compared.
For the lightest atoms, the choice is clear-cut,
beginning with H and He, where the density basis spans
the full set of occupied-occupied and occupied-virtual products,
we follow the pattern of eigenvalue distribution for all other atoms,
also keeping the overall accuracy $\mathcal{E}$ at nearly the same level.

\begingroup
\squeezetable
\begin{table*}[h]
\caption{\label{tab:n}Sizes\textsuperscript{a} of atomic wavefunction and density basis sets.}
\begin{ruledtabular}
\begin{tabular}{llrrrrrrrrcrrrrrrrrrrrllrrrrrrrrcrrrrrrrrrrr}
atoms & set & \multicolumn{8}{c}{wavefunction} && \multicolumn{11}{c}{density} &
atoms & set & \multicolumn{8}{c}{wavefunction} && \multicolumn{11}{c}{density} \\
 \cline{3-10}  \cline{12-22}
 \cline{25-32} \cline{34-44}
 H  -- He &\texttt{L1  }&  2& 1&  &  &  &  &  &  &&  2& 1&  &  &  &  &  &  &  &  &   & In -- Xe &\texttt{L1  }&  6& 5& 3&  &  &  &  &  && 11&10& 9& 6& 3&  &  &  &  &  &   \\
          &\texttt{L2  }&  3& 2& 1&  &  &  &  &  &&  3& 2& 1&  &  &  &  &  &  &  &   &          &\texttt{L2  }&  7& 6& 4& 1&  &  &  &  && 12&11&10& 7& 5&  &  &  &  &  &   \\
          &\texttt{L3  }&  4& 3& 2& 1&  &  &  &  &&  4& 3& 2& 1&  &  &  &  &  &  &   &          &\texttt{L3  }&  8& 7& 5& 2& 1&  &  &  && 12&11&11& 8& 6& 1&  &  &  &  &   \\
          &\texttt{L4  }&  5& 4& 3& 2& 1&  &  &  &&  5& 4& 3& 2& 1&  &  &  &  &  &   &          &\texttt{L4  }&  9& 8& 6& 3& 2& 1&  &  && 13&12&12& 9& 7& 3& 1&  &  &  &   \\
          &\texttt{L1a }&  3& 2&  &  &  &  &  &  &&  3& 2&  &  &  &  &  &  &  &  &   &          &\texttt{L1a }&  7& 6& 4&  &  &  &  &  && 12&11&10& 7& 3&  &  &  &  &  &   \\
          &\texttt{L2a }&  4& 3& 2&  &  &  &  &  &&  4& 3& 2&  &  &  &  &  &  &  &   &          &\texttt{L2a }&  8& 7& 5& 2&  &  &  &  && 13&12&11& 8& 6&  &  &  &  &  &   \\
          &\texttt{L3a }&  5& 4& 3& 2&  &  &  &  &&  5& 4& 3& 2&  &  &  &  &  &  &   &          &\texttt{L3a }&  9& 8& 6& 3& 2&  &  &  && 13&12&12& 9& 7& 2&  &  &  &  &   \\
          &\texttt{L4a }&  6& 5& 4& 3& 2&  &  &  &&  6& 5& 4& 3& 2&  &  &  &  &  &   &          &\texttt{L4a }& 10& 9& 7& 4& 3& 2&  &  && 14&13&13&10& 8& 4& 2&  &  &  &   \\
 Li -- Ne &\texttt{L11 }&  4& 3& 1&  &  &  &  &  &&  5& 4& 2& 1&  &  &  &  &  &  &   &          &\texttt{L11a}&  8& 7& 5& 1&  &  &  &  && 12&11&10& 8& 4& 1&  &  &  &  &   \\
          &\texttt{L22 }&  6& 5& 3& 1&  &  &  &  &&  6& 5& 4& 2& 1&  &  &  &  &  &   &          &\texttt{L22a}& 10& 9& 7& 4& 1&  &  &  && 13&12&12& 9& 7& 2& 1&  &  &  &   \\
          &\texttt{L33 }&  8& 7& 5& 3& 1&  &  &  &&  8& 7& 6& 4& 2& 1&  &  &  &  &   &          &\texttt{L33a}& 12&11& 9& 6& 4& 1&  &  && 14&13&13&11& 9& 6& 2& 1&  &  &   \\
          &\texttt{L44 }& 10& 9& 7& 5& 3& 1&  &  && 10& 9& 8& 6& 4& 2& 1&  &  &  &   &          &\texttt{L44a}& 14&13&11& 8& 6& 4& 1&  && 16&15&15&13&11& 8& 5& 2& 1&  &   \\
 B  -- Ne &\texttt{L1  }&  3& 2& 1&  &  &  &  &  &&  4& 3& 2& 1&  &  &  &  &  &  &   & Cs       &\texttt{L11 }&  8& 7& 4&  &  &  &  &  && 14&13&12& 7& 3&  &  &  &  &  &   \\
          &\texttt{L2  }&  4& 3& 2& 1&  &  &  &  &&  5& 4& 3& 2& 1&  &  &  &  &  &   &          &\texttt{L22 }& 10& 9& 6& 2&  &  &  &  && 15&14&14& 9& 6&  &  &  &  &  &   \\
          &\texttt{L3  }&  5& 4& 3& 2& 1&  &  &  &&  6& 5& 4& 3& 2& 1&  &  &  &  &   &          &\texttt{L33 }& 12&11& 8& 4& 2&  &  &  && 16&15&15&11& 8& 3&  &  &  &  &   \\
          &\texttt{L4  }&  6& 5& 4& 3& 2& 1&  &  &&  7& 6& 5& 4& 3& 2& 1&  &  &  &   &          &\texttt{L44 }& 14&13&10& 6& 4& 2&  &  && 17&16&16&12& 9& 4& 2&  &  &  &   \\
          &\texttt{L1a }&  4& 3& 2&  &  &  &  &  &&  5& 4& 3& 2&  &  &  &  &  &  &   &          &\texttt{L111}&  9& 8& 5& 1&  &  &  &  && 14&13&12& 8& 4& 1&  &  &  &  &   \\
          &\texttt{L2a }&  5& 4& 3& 2&  &  &  &  &&  6& 5& 4& 3& 2&  &  &  &  &  &   &          &\texttt{L222}& 12&11& 8& 4& 1&  &  &  && 15&14&14&10& 7& 3& 1&  &  &  &   \\
          &\texttt{L3a }&  6& 5& 4& 3& 2&  &  &  &&  7& 6& 5& 4& 3& 2&  &  &  &  &   &          &\texttt{L333}& 15&14&11& 7& 4& 1&  &  && 17&16&16&12&10& 6& 2& 1&  &  &   \\
          &\texttt{L4a }&  7& 6& 5& 4& 3& 2&  &  &&  8& 7& 6& 5& 4& 3& 2&  &  &  &   &          &\texttt{L444}& 18&17&14&10& 7& 4& 1&  && 18&18&18&15&12& 9& 5& 2& 1&  &   \\
          &\texttt{L11a}&  5& 4& 2&  &  &  &  &  &&  6& 5& 3& 2&  &  &  &  &  &  &   & Ba       &\texttt{L11 }&  8& 7& 5&  &  &  &  &  && 14&13&12& 9& 6&  &  &  &  &  &   \\
          &\texttt{L22a}&  7& 6& 4& 2&  &  &  &  &&  7& 6& 5& 3& 2&  &  &  &  &  &   &          &\texttt{L22 }& 10& 9& 7& 2&  &  &  &  && 15&14&14&10& 8& 2&  &  &  &  &   \\
          &\texttt{L33a}&  9& 8& 6& 4& 2&  &  &  &&  9& 8& 7& 5& 3& 2&  &  &  &  &   &          &\texttt{L33 }& 12&11& 9& 4& 2&  &  &  && 16&15&15&11& 9& 4& 1&  &  &  &   \\
          &\texttt{L44a}& 11&10& 8& 6& 4& 2&  &  && 11&10& 9& 7& 5& 3& 2&  &  &  &   &          &\texttt{L44 }& 14&13&11& 6& 4& 2&  &  && 17&16&16&12&10& 6& 4& 1&  &  &   \\
 Na -- Ar &\texttt{L11 }&  5& 4& 2&  &  &  &  &  &&  7& 6& 5& 2&  &  &  &  &  &  &   &          &\texttt{L111}&  9& 8& 6& 1&  &  &  &  && 14&13&12& 9& 7& 1&  &  &  &  &   \\
          &\texttt{L22 }&  7& 6& 4& 2&  &  &  &  &&  8& 7& 6& 4& 2&  &  &  &  &  &   &          &\texttt{L222}& 12&11& 9& 4& 1&  &  &  && 15&14&14&11& 9& 4& 1&  &  &  &   \\
          &\texttt{L33 }&  9& 8& 6& 4& 2&  &  &  &&  9& 9& 8& 6& 4& 2&  &  &  &  &   &          &\texttt{L333}& 15&14&12& 7& 4& 1&  &  && 17&16&16&13&11& 7& 4& 1&  &  &   \\
          &\texttt{L44 }& 11&10& 8& 6& 4& 2&  &  && 11&11&10& 8& 6& 4& 2&  &  &  &   &          &\texttt{L444}& 18&17&15&10& 7& 4& 1&  && 18&18&18&15&13&11& 7& 4& 1&  &   \\
 Al -- Ar &\texttt{L1  }&  4& 3& 1&  &  &  &  &  &&  6& 5& 4& 1&  &  &  &  &  &  &   & La       &\texttt{L11 }&  8& 7& 5& 1&  &  &  &  && 14&13&12& 9& 6& 1&  &  &  &  &   \\
          &\texttt{L2  }&  5& 4& 2& 1&  &  &  &  &&  7& 6& 5& 2& 1&  &  &  &  &  &   &          &\texttt{L22 }& 10& 9& 7& 3& 1&  &  &  && 15&14&14&10& 8& 3& 1&  &  &  &   \\
          &\texttt{L3  }&  6& 5& 3& 2& 1&  &  &  &&  8& 7& 6& 3& 2& 1&  &  &  &  &   &          &\texttt{L33 }& 12&11& 9& 5& 3& 1&  &  && 16&15&15&11& 9& 6& 2& 1&  &  &   \\
          &\texttt{L4  }&  7& 6& 4& 3& 2& 1&  &  &&  9& 8& 7& 4& 3& 2& 1&  &  &  &   &          &\texttt{L44 }& 14&13&11& 7& 5& 3& 1&  && 17&16&16&12&10& 7& 4& 2& 1&  &   \\
          &\texttt{L1a }&  5& 4& 2&  &  &  &  &  &&  7& 6& 5& 2&  &  &  &  &  &  &   &          &\texttt{L111}&  9& 8& 6& 2&  &  &  &  && 14&13&12& 9& 7& 2&  &  &  &  &   \\
          &\texttt{L2a }&  6& 5& 3& 2&  &  &  &  &&  8& 7& 6& 3& 2&  &  &  &  &  &   &          &\texttt{L222}& 12&11& 9& 5& 2&  &  &  && 15&14&14&11& 9& 5& 2&  &  &  &   \\
          &\texttt{L3a }&  7& 6& 4& 3& 2&  &  &  &&  9& 8& 7& 4& 3& 2&  &  &  &  &   &          &\texttt{L333}& 15&14&12& 8& 5& 2&  &  && 17&16&16&13&11& 8& 4& 2&  &  &   \\
          &\texttt{L4a }&  8& 7& 5& 4& 3& 2&  &  && 10& 9& 8& 5& 4& 3& 2&  &  &  &   &          &\texttt{L444}& 18&17&15&11& 8& 5& 2&  && 18&18&18&15&13&11& 7& 4& 2&  &   \\
          &\texttt{L11a}&  6& 5& 3&  &  &  &  &  &&  8& 7& 6& 3&  &  &  &  &  &  &   & Ce -- Rn &\texttt{L111}&  9& 8& 6& 3& 1&  &  &  && 14&13&13&10& 8& 5& 2& 1&  &  &   \\
          &\texttt{L22a}&  8& 7& 5& 3&  &  &  &  &&  9& 8& 7& 5& 3&  &  &  &  &  &   &          &\texttt{L222}& 12&11& 9& 6& 3& 1&  &  && 15&14&14&12&10& 7& 4& 2& 1&  &   \\
          &\texttt{L33a}& 10& 9& 7& 5& 3&  &  &  && 10&10& 9& 7& 5& 3&  &  &  &  &   &          &\texttt{L333}& 15&14&12& 9& 6& 3& 1&  && 17&16&16&13&11& 9& 6& 4& 2& 1&   \\
          &\texttt{L44a}& 12&11& 9& 7& 5& 3&  &  && 12&12&11& 9& 7& 5& 3&  &  &  &   &          &\texttt{L444}& 18&17&15&12& 9& 6& 3& 1&& 18&18&18&16&14&12& 9& 6& 4& 2& 1 \\
 K        &\texttt{L11 }&  6& 5& 2&  &  &  &  &  &&  9& 8& 7& 2&  &  &  &  &  &  &   & Lu -- Rn &\texttt{L11 }&  8& 7& 5& 2&  &  &  &  && 14&13&13&10& 7& 4& 1&  &  &  &   \\
          &\texttt{L22 }&  8& 7& 4& 2&  &  &  &  && 10& 9& 8& 4& 2&  &  &  &  &  &   &          &\texttt{L22 }& 10& 9& 7& 4& 1&  &  &  && 15&14&14&11& 9& 6& 3&  &  &  &   \\
          &\texttt{L33 }& 10& 9& 6& 4& 2&  &  &  && 11&10& 9& 6& 4& 2&  &  &  &  &   &          &\texttt{L33 }& 12&11& 9& 6& 3& 1&  &  && 16&15&15&12&10& 7& 4& 1&  &  &   \\
          &\texttt{L44 }& 12&11& 8& 6& 4& 2&  &  && 13&12&11& 9& 6& 4& 2&  &  &  &   &          &\texttt{L44 }& 14&13&11& 8& 5& 3& 1&  && 17&16&16&13&11& 9& 6& 2& 1&  &   \\
 Ca       &\texttt{L11 }&  6& 5& 3&  &  &  &  &  &&  9& 8& 7& 5& 2&  &  &  &  &  &   & Tl -- Rn &\texttt{L1  }&  7& 6& 4& 1&  &  &  &  && 14&13&13& 9& 6& 3& 1&  &  &  &   \\
          &\texttt{L22 }&  8& 7& 5& 2&  &  &  &  && 10& 9& 8& 6& 4& 1&  &  &  &  &   &          &\texttt{L2  }&  8& 7& 5& 2&  &  &  &  && 15&14&14&11& 8& 4& 1&  &  &  &   \\
          &\texttt{L33 }& 10& 9& 7& 4& 2&  &  &  && 11&10&10& 8& 6& 4& 1&  &  &  &   &          &\texttt{L3  }&  9& 8& 6& 3& 1&  &  &  && 15&14&14&11& 9& 5& 1&  &  &  &   \\
          &\texttt{L44 }& 12&11& 9& 6& 4& 2&  &  && 13&12&12&10& 8& 6& 4& 1&  &  &   &          &\texttt{L4  }& 10& 9& 7& 4& 2& 1&  &  && 16&15&15&12&10& 6& 3&  &  &  &   \\
 Sc -- Kr &\texttt{L11 }&  6& 5& 3& 1&  &  &  &  &&  9& 8& 7& 5& 2& 1&  &  &  &  &   & Fr       &\texttt{L11 }&  9& 8& 5& 1&  &  &  &  && 16&15&15&11& 7& 3& 1&  &  &  &   \\
          &\texttt{L22 }&  8& 7& 5& 3& 1&  &  &  && 10& 9& 8& 6& 4& 2& 1&  &  &  &   &          &\texttt{L22 }& 11&10& 7& 3&  &  &  &  && 18&17&17&13& 9& 4& 1&  &  &  &   \\
          &\texttt{L33 }& 10& 9& 7& 5& 3& 1&  &  && 11&10&10& 8& 6& 4& 2& 1&  &  &   &          &\texttt{L33 }& 13&12& 9& 5& 2&  &  &  && 19&18&18&14&11& 6& 1&  &  &  &   \\
          &\texttt{L44 }& 12&11& 9& 7& 5& 3& 1&  && 13&12&12&10& 8& 6& 4& 2& 1&  &   &          &\texttt{L44 }& 15&14&11& 7& 4& 2&  &  && 20&20&20&16&13& 8& 4&  &  &  &   \\
 Ga -- Kr &\texttt{L1  }&  5& 4& 2&  &  &  &  &  &&  9& 8& 7& 4& 1&  &  &  &  &  &   &          &\texttt{L111}& 10& 9& 6& 2&  &  &  &  && 16&15&15&11& 7& 4& 1&  &  &  &   \\
          &\texttt{L2  }&  6& 5& 3& 1&  &  &  &  && 10& 9& 8& 5& 3&  &  &  &  &  &   &          &\texttt{L222}& 13&12& 9& 5& 1&  &  &  && 18&17&17&13&10& 6& 3&  &  &  &   \\
          &\texttt{L3  }&  7& 6& 4& 2& 1&  &  &  && 10& 9& 9& 6& 4& 1&  &  &  &  &   &          &\texttt{L333}& 16&15&12& 8& 4& 1&  &  && 19&18&18&15&12& 8& 4& 1&  &  &   \\
          &\texttt{L4  }&  8& 7& 5& 3& 2& 1&  &  && 11&10&10& 7& 5& 3& 1&  &  &  &   &          &\texttt{L444}& 19&18&15&11& 7& 4& 1&  && 20&20&20&17&15&11& 8& 3& 1&  &   \\
          &\texttt{L1a }&  6& 5& 3&  &  &  &  &  && 10& 9& 8& 5& 1&  &  &  &  &  &   & Ra       &\texttt{L11 }&  9& 8& 6& 1&  &  &  &  && 16&15&15&12& 9& 3& 1&  &  &  &   \\
          &\texttt{L2a }&  7& 6& 4& 2&  &  &  &  && 11&10& 9& 6& 4&  &  &  &  &  &   &          &\texttt{L22 }& 11&10& 8& 3&  &  &  &  && 18&17&17&14&11& 6& 1&  &  &  &   \\
          &\texttt{L3a }&  8& 7& 5& 3& 2&  &  &  && 11&10&10& 7& 5& 2&  &  &  &  &   &          &\texttt{L33 }& 13&12&10& 5& 2&  &  &  && 19&18&18&15&13& 8& 4&  &  &  &   \\
          &\texttt{L4a }&  9& 8& 6& 4& 3& 2&  &  && 12&11&11& 8& 6& 4& 2&  &  &  &   &          &\texttt{L44 }& 15&14&12& 7& 4& 2&  &  && 20&20&20&16&14& 9& 5& 1&  &  &   \\
          &\texttt{L11a}&  7& 6& 4& 1&  &  &  &  && 10& 9& 8& 6& 2& 1&  &  &  &  &   &          &\texttt{L111}& 10& 9& 7& 2&  &  &  &  && 16&15&15&12&10& 4& 1&  &  &  &   \\
          &\texttt{L22a}&  9& 8& 6& 4& 1&  &  &  && 11&10& 9& 7& 5& 2& 1&  &  &  &   &          &\texttt{L222}& 13&12&10& 5& 1&  &  &  && 18&17&17&14&12& 7& 3&  &  &  &   \\
          &\texttt{L33a}& 11&10& 8& 6& 4& 1&  &  && 12&11&11& 9& 7& 5& 2& 1&  &  &   &          &\texttt{L333}& 16&15&13& 8& 4& 1&  &  && 19&18&18&16&14&10& 6& 1&  &  &   \\
          &\texttt{L44a}& 13&12&10& 8& 6& 4& 1&  && 14&13&13&11& 9& 7& 5& 2& 1&  &   &          &\texttt{L444}& 19&18&16&11& 7& 4& 1&  && 20&20&20&17&15&12& 9& 4& 1&  &   \\
 Rb       &\texttt{L11 }&  7& 6& 3&  &  &  &  &  && 11&10& 9& 5& 1&  &  &  &  &  &   & Ac       &\texttt{L11 }&  9& 8& 6& 2&  &  &  &  && 16&15&15&12& 9& 4& 1&  &  &  &   \\
          &\texttt{L22 }&  9& 8& 5& 2&  &  &  &  && 12&11&11& 7& 3&  &  &  &  &  &   &          &\texttt{L22 }& 11&10& 8& 4& 1&  &  &  && 18&17&17&14&11& 7& 2&  &  &  &   \\
          &\texttt{L33 }& 11&10& 7& 4& 2&  &  &  && 13&12&12& 9& 6& 2&  &  &  &  &   &          &\texttt{L33 }& 13&12&10& 6& 3& 1&  &  && 19&18&18&15&13& 8& 5& 1&  &  &   \\
          &\texttt{L44 }& 13&12& 9& 6& 4& 2&  &  && 15&14&14&11& 8& 4& 2&  &  &  &   &          &\texttt{L44 }& 15&14&12& 8& 5& 3& 1&  && 20&20&20&16&14&10& 7& 2& 1&  &   \\
 Sr       &\texttt{L11 }&  7& 6& 4&  &  &  &  &  && 11&10& 9& 7& 4&  &  &  &  &  &   &          &\texttt{L111}& 10& 9& 7& 3&  &  &  &  && 16&15&15&12&10& 5& 1&  &  &  &   \\
          &\texttt{L22 }&  9& 8& 6& 2&  &  &  &  && 12&11&11& 8& 6& 1&  &  &  &  &   &          &\texttt{L222}& 13&12&10& 6& 2&  &  &  && 18&17&17&14&12& 8& 4&  &  &  &   \\
          &\texttt{L33 }& 11&10& 8& 4& 2&  &  &  && 13&12&12&10& 8& 4& 1&  &  &  &   &          &\texttt{L333}& 16&15&13& 9& 5& 2&  &  && 19&18&18&16&14&11& 7& 2&  &  &   \\
          &\texttt{L44 }& 13&12&10& 6& 4& 2&  &  && 15&14&14&12&10& 6& 4& 1&  &  &   &          &\texttt{L444}& 19&18&16&12& 8& 5& 2&  && 20&20&20&17&15&12&10& 5& 2&  &   \\
 Y  -- Xe &\texttt{L11 }&  7& 6& 4& 1&  &  &  &  && 11&10& 9& 7& 4& 1&  &  &  &  &   & Th -- No &\texttt{L111}& 10& 9& 7& 4& 1&  &  &  && 16&15&15&12&10& 7& 4& 1&  &  &   \\
          &\texttt{L22 }&  9& 8& 6& 3& 1&  &  &  && 12&11&11& 8& 6& 2& 1&  &  &  &   &          &\texttt{L222}& 13&12&10& 7& 3& 1&  &  && 18&17&17&14&12& 9& 6& 2& 1&  &   \\
          &\texttt{L33 }& 11&10& 8& 5& 3& 1&  &  && 13&12&12&10& 8& 5& 2& 1&  &  &   &          &\texttt{L333}& 16&15&13&10& 6& 3& 1&  && 19&18&18&16&14&11& 8& 4& 2& 1&   \\
          &\texttt{L44 }& 13&12&10& 7& 5& 3& 1&  && 15&14&14&12&10& 7& 4& 2& 1&  &   &          &\texttt{L444}& 19&18&16&13& 9& 6& 3& 1&& 20&20&20&18&16&13&11& 8& 4& 2& 1 \\
\end{tabular}
\begin{flushleft}
\textsuperscript{a}Number of radial functions for angular momenta $l=0,\dots,l_\mathrm{max}$.
\end{flushleft}
\end{ruledtabular}
\end{table*}
\endgroup

Another choice to be made is the number of underlying primitive functions ---
starting from some high number for each angular symmetry,
we bring it down until the error rises to no more than twice the starting value.
For the lower angular momenta, more primitives need to be contracted,
and even more so for the heavier elements,
while the smallest number is found enough for the higher.

We tabulate the results, for all 102 elements, in the
\href{ftp://ftp.aip.org/epaps/journ_chem_phys/}{supplementary material}
so that the files can be concatenated with those from our other work~\cite{L19a}
to get a full package of atomic basis sets for theoretical chemistry.
(The non-relativistic point-nucleus analogs for the first 36 elements
are also included.)

\begingroup
\squeezetable
\begin{table}\caption{\label{tab:m}Tests of molecular integral approximation.}
\begin{ruledtabular}
\begin{tabular}{llrrrr}
molecule & set &
 \multicolumn{1}{c}{$E_\mathrm{HF}$} &
 \multicolumn{1}{c}{$E_\mathrm{MP2}$} &
 \multicolumn{1}{c}{$\delta E_\mathrm{HF}$\textsuperscript{a}} &
 \multicolumn{1}{c}{$\delta E_\mathrm{MP2}$\textsuperscript{a}} \\
H$_3$N
 & \texttt{L1\_\tiny{}3}   & -56.257006 & -0.198373 & -0.000060 & 0.000797 \\
 & \texttt{L1a\_\tiny{}3}  & -56.264521 & -0.214041 & -0.000064 & 0.000219 \\
 & \texttt{L2\_\tiny{}3}   & -56.264915 & -0.239455 & -0.000012 & 0.000393 \\
 & \texttt{L2a\_\tiny{}3}  & -56.268333 & -0.245274 & -0.000029 & 0.000183 \\
 & \texttt{L3\_\tiny{}3}   & -56.267775 & -0.253086 & -0.000032 & 0.000198 \\
 & \texttt{L3a\_\tiny{}3}  & -56.269261 & -0.255372 & -0.000009 & 0.000122 \\
 & \texttt{L33a\_\tiny{}3} & -56.269479 & -0.311461 & -0.000016 & 0.000136 \\
 & \texttt{L4\_\tiny{}3}   & -56.268910 & -0.258385 & -0.000017 & 0.000110 \\
 & \texttt{L4a\_\tiny{}3}  & -56.269465 & -0.259446 & -0.000011 & 0.000070 \\
\\
H$_2$O
 & \texttt{L1\_\tiny{}3}   & -76.134433 & -0.218305 &  0.000185 & 0.000628 \\
 & \texttt{L1a\_\tiny{}3}  & -76.141327 & -0.238776 &  0.000045 & 0.000097 \\
 & \texttt{L2\_\tiny{}3}   & -76.143573 & -0.267992 &  0.000024 & 0.000246 \\
 & \texttt{L2a\_\tiny{}3}  & -76.145853 & -0.275971 & -0.000028 & 0.000122 \\
 & \texttt{L3\_\tiny{}3}   & -76.145706 & -0.285628 & -0.000031 & 0.000150 \\
 & \texttt{L3a\_\tiny{}3}  & -76.146742 & -0.288898 & -0.000010 & 0.000099 \\
 & \texttt{L33a\_\tiny{}3} & -76.146912 & -0.348174 & -0.000018 & 0.000105 \\
 & \texttt{L4\_\tiny{}3}   & -76.146537 & -0.292598 & -0.000032 & 0.000092 \\
 & \texttt{L4a\_\tiny{}3}  & -76.146920 & -0.294123 & -0.000014 & 0.000056 \\
 & &
 \multicolumn{1}{c}{$\Delta E_\mathrm{HF}$\textsuperscript{b}} &
 \multicolumn{1}{c}{$\Delta E_\mathrm{MP2}$\textsuperscript{b}} &
 \multicolumn{1}{c}{$\delta\Delta E_\mathrm{HF}$} &
 \multicolumn{1}{c}{$\delta\Delta E_\mathrm{MP2}$} \\
H$_3$N{\dots}H$_2$O
 & \texttt{L1\_\tiny{}3}   & -0.008680 & -0.003151 &  0.000012 & -0.000024 \\
 & \texttt{L1a\_\tiny{}3}  & -0.007218 & -0.004258 &  0.000016 & -0.000027 \\
 & \texttt{L2\_\tiny{}3}   & -0.007428 & -0.003663 &  0.000020 & -0.000011 \\
 & \texttt{L2a\_\tiny{}3}  & -0.006526 & -0.004119 &  0.000009 &  0.000002 \\
 & \texttt{L3\_\tiny{}3}   & -0.006879 & -0.003806 &  0.000006 & -0.000013 \\
 & \texttt{L3a\_\tiny{}3}  & -0.006484 & -0.004023 & -0.000003 &  0.000001 \\
 & \texttt{L33a\_\tiny{}3} & -0.006433 & -0.004099 & -0.000005 & -0.000002 \\
 & \texttt{L4\_\tiny{}3}   & -0.006614 & -0.003909 &  0.000003 & -0.000007 \\
 & \texttt{L4a\_\tiny{}3}  & -0.006441 & -0.003997 &  0.000001 &  0.000002 \\
\end{tabular}
\begin{flushleft}
\textsuperscript{a}Energy error $\delta E = \tilde{E} - E$, au. \\
\textsuperscript{b}Binding energy
$\Delta E = E(\mathrm{complex}) - E(\mathrm{mol 1}) - E(\mathrm{mol 2})$, au.
\end{flushleft}
\end{ruledtabular}
\end{table}
\endgroup

\begingroup
\squeezetable
\begin{table}\caption{\label{tab:t}Further tests across the periodic table.}
\begin{ruledtabular}
\begin{tabular}{lrrrrrr}
set &
 \multicolumn{1}{c}{$\delta E_\mathrm{HF}$} &
 \multicolumn{1}{c}{$\delta E_\mathrm{MP2}$} &
 \multicolumn{1}{c}{$\delta E_\mathrm{HF}$} &
 \multicolumn{1}{c}{$\delta E_\mathrm{MP2}$} &
 \multicolumn{1}{c}{$\delta E_\mathrm{HF}$} &
 \multicolumn{1}{c}{$\delta E_\mathrm{MP2}$} \\
&
 \multicolumn{2}{c}{Li$_2$} &
 \multicolumn{2}{c}{Na$_2$} &
 \multicolumn{2}{c}{K$_2$} \\
 \texttt{L1\_\tiny{}3} &-0.000030 & 0.000112 & 0.000317 &-0.000415 & 0.000416 & 0.001084 \\
 \texttt{L2\_\tiny{}3} & 0.000023 & 0.000106 & 0.000101 & 0.000299 & 0.000074 & 0.001385 \\
 \texttt{L3\_\tiny{}3} &-0.000009 & 0.000144 &-0.000056 & 0.000288 &-0.000133 & 0.001054 \\
 \texttt{L4\_\tiny{}3} &-0.000004 & 0.000067 &-0.000061 & 0.000191 &-0.000145 & 0.000229 \\
&
 \multicolumn{2}{c}{Rb$_2$} &
 \multicolumn{2}{c}{Cs$_2$} &
 \multicolumn{2}{c}{Fr$_2$} \\
 \texttt{L1\_\tiny{}3} & 0.000037 & 0.001039 & 0.000405 & 0.001022 & 0.000585 & 0.000210 \\
 \texttt{L2\_\tiny{}3} &-0.000004 & 0.001299 & 0.000091 & 0.000438 &-0.000313 & 0.000094 \\
 \texttt{L3\_\tiny{}3} &-0.000427 & 0.000538 &-0.001260 & 0.000428 &          &          \\
 \texttt{L4\_\tiny{}3} &-0.000547 & 0.000219 &-0.001402 & 0.001513 &          &          \\
&
 \multicolumn{2}{c}{N$_2$} &
 \multicolumn{2}{c}{P$_2$} &
 \multicolumn{2}{c}{As$_2$} \\
 \texttt{L1\_\tiny{}3} & 0.001030 & 0.000408 & 0.001347 &-0.000074 & 0.000609 & 0.000102 \\
 \texttt{L2\_\tiny{}3} & 0.000096 & 0.000102 & 0.000468 & 0.000173 &-0.000109 & 0.000057 \\
 \texttt{L3\_\tiny{}3} & 0.000009 & 0.000095 & 0.000125 & 0.000200 &-0.000045 & 0.000169 \\
 \texttt{L4\_\tiny{}3} &-0.000013 & 0.000072 &-0.000036 & 0.000111 &-0.000141 & 0.000080 \\
&
 \multicolumn{2}{c}{Sb$_2$} &
 \multicolumn{2}{c}{Bi$_2$} \\
 \texttt{L1\_\tiny{}3} & 0.000901 & 0.000264 & 0.002491 & 0.000126  \\
 \texttt{L2\_\tiny{}3} & 0.000139 & 0.000119 & 0.000458 & 0.000008  \\
 \texttt{L3\_\tiny{}3} & 0.000266 & 0.000221 & 0.000188 & 0.000082  \\
 \texttt{L4\_\tiny{}3} &-0.000094 & 0.000098 &-0.000190 & 0.000040  \\
 &
 \multicolumn{2}{c}{F$_2$} &
 \multicolumn{2}{c}{Cl$_2$} &
 \multicolumn{2}{c}{Br$_2$} \\
 \texttt{L1\_\tiny{}3}  & 0.001375 & 0.001232 & 0.001482 & 0.000877 & 0.000104 & 0.000883 \\
 \texttt{L1a\_\tiny{}3} & 0.000332 & 0.000125 & 0.000856 & 0.000627 & 0.000307 & 0.000706 \\
 \texttt{L2\_\tiny{}3}  & 0.000274 & 0.000252 & 0.000462 & 0.000751 &-0.000053 & 0.000228 \\
 \texttt{L2a\_\tiny{}3} & 0.000053 & 0.000080 & 0.000287 & 0.000642 &-0.000061 & 0.000134 \\
 \texttt{L3\_\tiny{}3}  & 0.000050 & 0.000142 & 0.000087 & 0.000522 &-0.000052 & 0.000347 \\
 \texttt{L3a\_\tiny{}3} & 0.000020 & 0.000096 &-0.000034 & 0.000308 &-0.000008 & 0.000150 \\
 \texttt{L4\_\tiny{}3}  & 0.000011 & 0.000100 &-0.000008 & 0.000232 &-0.000012 & 0.000142 \\
 \texttt{L4a\_\tiny{}3} & 0.000000 & 0.000053 &-0.000067 & 0.000168 &-0.000138 & 0.000091 \\
 &
 \multicolumn{2}{c}{I$_2$} & & &
 \multicolumn{2}{c}{H$_2$} \\
 \texttt{L1\_\tiny{}3}  & 0.001001 & 0.000904 & & &-0.000068 & 0.000329 \\
 \texttt{L1a\_\tiny{}3} & 0.000693 & 0.000833 & & &-0.000029 & 0.000187 \\
 \texttt{L2\_\tiny{}3}  & 0.000193 & 0.000369 & & &-0.000007 & 0.000196 \\
 \texttt{L2a\_\tiny{}3} & 0.000102 & 0.000262 & & &-0.000003 & 0.000089 \\
 \texttt{L3\_\tiny{}3}  & 0.000262 & 0.000474 & & &-0.000002 & 0.000081 \\
 \texttt{L3a\_\tiny{}3} & 0.000057 & 0.000239 & & &-0.000002 & 0.000045 \\
 \texttt{L4\_\tiny{}3}  & 0.000023 & 0.000176 & & &-0.000001 & 0.000038 \\
 \texttt{L4a\_\tiny{}3} &-0.000097 & 0.000078 & & &-0.000000 & 0.000023 \\
\end{tabular}
\end{ruledtabular}
\end{table}
\endgroup

We have done many tests on molecules
for the accuracy of HF~\cite{H28,F30} and MP2~\cite{MP34} correlation
energy components of wavefunction theory,
and also for Coulomb and exchange-correlation~\cite{L97} components
within the generalized-gradient approximation~\cite{PBE96}
of density-functional~\cite{KS65} theory,
using the global fit with Coulomb metric to approximate the integrals
and comparing the results with those from the exact evaluation ---
the agreement is good,
we show in Table~\ref{tab:m}
the characteristic example of water-ammonia complex
as a small but meaningful fragment of chemical matter,
and in Table~\ref{tab:t} some further tests on simple diatomics
across the periodic table.
(We have found only one problem for the heaviest elements (after Xe)
with the biggest (\texttt{L3}, \texttt{L4}) basis sets ---
the many and large opposite-sign coefficients in the wavefunction basis
lead to a severe loss of precision in the two-electron integrals,
so that the IEEE-754~\cite{IEEE754} double precision (53 significand bits)
is no more enough to get any meaningful values ---
but this has nothing to do with the density-fitting.
There is a solution: the (two- and) three-index integrals for molecules
can be calculated using local prolate spheroidal coordinate systems
and approximation or interpolation functions with parameters
precomputed in higher precision --- this could be not only more stable numerically,
but also much faster --- but we do not study it here.)

\section{Conclusions}

We are looking forward to further applications
of our density-fitting basis sets,
they have already been (and will be) used in computational studies
to help organic chemists~\cite{VLFSZTZMB20,ONKHLK20}
(including ourselves~\cite{BL20})
understand~\cite{HM20a,HM20b,HM20c} reaction mechanisms and design new molecules.

\section{Data availability}

See \href{ftp://ftp.aip.org/epaps/journ_chem_phys/}{supplementary material}
for data files.
Both wavefunction and density-fitting basis sets
may one day be uploaded to an open database~\cite{PADGW19}
--- the negotiations have already begun.

\section{Supplementary Material}

Density-fitting basis sets for hydrogen through nobelium
are tabulated in a both human- and machine-readable form.

\begin{acknowledgments}
We kindly thank Eduard E. Levin~\cite{LKPBVNS20,LVN17}
for raising our awareness of the ``writer's block''~\cite{U74}.
Many thanks to the Editor for keeping the peer-review process
from becoming a kind~\cite{C06} of run-away Milgram~\cite{M63} experiment.
\end{acknowledgments}

\clearpage


%

\end{document}